\documentclass[twocolumn,showpacs,pra,groupedaddress,superscriptaddress,nofootinbib,longbibliography]{revtex4-1}

\usepackage{graphicx,amsmath,amsthm,amssymb,amsbsy,subfigure,hyperref,bbm,times,txfonts,float}
\usepackage{subfigure,hyperref,bbm,times}
\usepackage[T1]{fontenc}
\usepackage{braket}
\usepackage{epsfig}
\usepackage{color}
\usepackage{graphicx}
\usepackage{dcolumn}
\usepackage{bm}

\providecommand{\openone}{\leavevmode\hbox{\small1\kern-3.8pt\normalsize1}}

\usepackage{soul}

\hypersetup{
	colorlinks=true,
	linkcolor=blue,
}
\graphicspath{{figure/}}

\begin{document}
	
	
	\title{Remote sensing and  faithful quantum teleportation   through non-localized qubits}
	
\author{Hossein Rangani Jahromi}
	\email{h.ranganijahromi@jahromu.ac.ir}
	\affiliation{Physics Department, Faculty of Sciences, Jahrom University, P.B. 74135111, Jahrom, Iran}

	\date{\today}

\begin{abstract}
	One of the most important applications of quantum physics is quantum teleportation,  the possibility to
	transfer quantum states over arbitrary distances.
In this paper, we address the idea of remote sensing in a teleportation scenario with topological qubits more robust against noise. We also investigate the enhancement of quantum teleportation through non-local characteristics of the topological qubits.  In particular, we show that how  this nonlocal property, helps us to achieve near-perfect quantum teleportation  even with mixed
quantum states.   Considering the limitations
imposed by decoherence and the subsequent mixedness of the resource state, we find that our results may solve  important challenges in realizing faithful teleportation over long distances.

\end{abstract}


\maketitle


\section{Introduction \label{introduction}}
Quantum teleportation, describing the transfer
of an unknown quantum state over a long distance, constitutes a fundamental problem due to unavoidable
transportation losses in combination with the no-cloning
theorem \cite{wootters1982single}.
First highlighted by Bennett \textit{et al.} \cite{bennett1993teleporting}, it has since evolved
into an active and interesting area of research and is now recognized as an
significant  tool for many quantum protocols such as  measurement-based quantum computing \cite{raussendorf2001one}, quantum
repeaters \cite{sangouard2011quantum}, 
and fault-tolerant quantum computation \cite{gottesman1999demonstrating}. The experiments
have been first implemented  by photons \cite{bouwmeester1997experimental}, later with various
systems such as trapped ions \cite{riebe2004deterministic, barrett2004deterministic}, atomic ensembles \cite{sherson2006quantum},
as well as with high-frequency phonons \cite{hou2016quantum} and several
others \cite{pirandola2015advances,kumar2020experimental,PhysRevA.102.062208,langenfeld2021quantum}.

\par
High-precision parameter  estimation is a fundamental task
throughout science. Generally speaking, there are two different scenarios for quantum parameter estimation in the presence of noises \cite{holevo1978estimation,giovannettiScience,paris2009quantum,giovannetti2011advances,liu2019quantum,
	toth2014quantum,polinoReview,pirandolaReview,NatRev2019,RevModPhys.90.035005}. In both approaches, the most important goals is to investigate the optimized measurement strategy
such that as much information as possible about some parameter  is achieved. In the first scenario, a quantum probe, with known initial state, is transmitted through a
quantum channel, encoding  the parameter of interest into the quantum state of the system, and then the output state is measured to extract  an estimate from  measurement results.
However, in the other standard scenario, the information about the quantity of interest is initially
encoded into the  system state and then this information carrier is transmitted
through a quantum noisy channel; finally, the measurement process is implemented.

\par We investigate the first aforementioned scenario 
where the information about the strength of the coupling between a topological qubit and its environment, is encoded by a teleportation channel \cite{pirandola2015advances} into the state of the teleported qubits, leading to idea of remote sensing. This strategy is applied when our metrological  setup is not accessible at a special place and we need to estimate some unknown parameter at that location without moving the devices. Experimental setups for teleportation-based quantum information processing with Majorana zero modes have been proposed in \cite{vijay2016teleportation}.

If the probes are classically correlated and noninteracting, as a consequence of the central limit theorem,
the mean-squared error of the estimate decreases as $ 1/N $,
in which $ N $ denotes the number of probes. This best
scaling achievable through a classical probe is known as the
standard quantum limit. Quantum metrology
aims to enhance estimation by exploiting quantum correlations between probes \cite{brask2015improved}. In the absence of decoherence, it is well known that
quantum resources allow for a quadratic improvement in
precision over the standard quantum limit.
However, in realistic evolution the presence of decoherence effects is 
unavoidable,  and hence there is currently much effort to investigate
exactly when and how quantum resources such as entanglement allow
estimation to be improved in the presence of noises \cite{wang2018entanglement,braun2018quantum,liu2021distributed,Jahromi2020,rangani2019weak}. We exploit this scenario through entangled teleported qubits realized by \textit{Majorana modes} for \textit{remote} quantum sensing of their couplings with the \textit{input} environment. Estimation  the system-environment coupling strength helps us to determine whether we can apply specific approximations in the theoretical model or not \cite{hu1992quantum,boyanovsky2005particle}. More importantly, the strength of the coupling is a significant parameter to control the decoherence effects \cite{ho2014decoherence,liu2013anomalous,wu2014quenched}. Moreover, in many-body quantum
systems, changing the coupling constant may drive the system
into different phases \cite{invernizzi2008optimal}. These reasons motivate us to estimate the  coupling constant of an interacting  Hamiltonian  not corresponding
to any observable of the system, and hence it cannot be measured directly.

\par
It has been shown that the topological quantum computation \cite{freedman2003topological}  is  one of the most exciting approaches to
constructing a fault-tolerant quantum computer  \cite{nayak2008non}. Particularly, there are novel kinds of topologically ordered states, such as topological
insulators and superconductors \cite{fu2008superconducting,hasan2010colloquium,qi2011topological},  easy to realize physically.  The most interesting
excitations for  these important systems are the Majorana modes, localized on topological defects,  which obey the non-Abelian
anyonic statistics \cite{wilczek2009majorana,arovas1984fractional,ivanov2001non}.
\par
One of the simplest scenarios for realization of Majorana modes  are those appearing at   the edges of the Kitaev's
 spineless p-wave superconductor chain model \cite{kitaev2001unpaired,sau2010generic,alicea2010majorana,thakurathi2014majorana} where  two far separated endpoint Majorana modes can compose a topological qubit.
  The most significant characteristic of the topological qubit is its  non-locality, since the two Majorana modes are far separated. This non-local property makes topological qubits interact  with the environment more unpredictably than the usual local qubits \cite{ho2014decoherence}. 
  Motivated by this, we study the implementation of quantum information tasks usually investigated for local qubits, such as quantum teleportation and quantum parameter estimation,  through non-local qubits.

\par 
A continuous process is called  Markovian if,
starting from any initial state, its dynamics can be determined unambiguously from the initial state. Non-Markovianity \cite{rivas2014quantum,breuer2012foundations} is inherently
connected to the two-way exchange of information between
the system and the environment; a Markovian description of dynamics is legitimate, even if only as an approximation, whenever the
observed time scale of the evolution is much larger than
the correlation time characterizing the interaction between system and its environment. Non-Markovianity is a complicated
phenomenon affecting the system both in its informational and dynamical features.
For a recent review of the witnesses and approaches to characterize non-Markovianity we refer to \cite{li2018concepts}.

\par In this paper  we consider remote  sensing  through quantum teleportation implemented by  non-local Majorana modes  realizing  two topological qubits independently  coupled to  non-Markovian  Ohmic-like reservoirs. We show that how environmental control parameters, i.e., cutoff frequency, Ohmicity parameter, and the coupling strength, applied at the destination of teleportation, affect  the remote sensing, the quality of teleportation  and quantum correlations between teleported non-local qubits. In particular, the quantum control  to achieve near-perfect teleportation is discussed.

\par This paper is organized as follows: In Section \ref{PRELIMINARIES}, we
present a brief review of the standard quantum teleportation, quantum metrology, measure of quantum resources and Hilbert-Schmidt speed. The model and its non-Markovian characteristics are introduced in Section \ref{Model}. Moreover, the scenario of teleportation and remote  sensing are discussed in Sec. \ref{Tel}.  
Finally in Section \ref{cunclusion}, the main results are summarized.

\section{PRELIMINARIES}\label{PRELIMINARIES}
In this section we review the most important concepts discussed in this paper. 

\subsubsection{Standard quantum teleportation}
The main idea of quantum teleportation is transferring quantum information from the sender (Alice) to the receiver (Bob) through  entangled pairs.  In the standard  protocol \cite{bowen2001teleportation},  the teleportation is realized by a two-qubit mixed state 
$ \rho_{\text{res}} $,
playing the role of the resource, and is modeled by a generalized depolarizing channel $\Lambda_{\text{res}}$, acting on an input state $\rho_{\text{in}}$ which is  the single-qubit state to be teleported, i.e.,
\begin{eqnarray}\label{a1}
	\nonumber\rho_{\text{out}}=\Lambda_{\text{res}}\left( \rho_{\text{in}}\right),~~~~~~~~~~~~~\\
	=\sum^{3}_{i=0}\text{Tr}\left[\mathcal{B}_{i}\rho_{\text{\text{res}}} \right]\sigma_{i}\rho_{\text{in}}\sigma_{i}, 
\end{eqnarray}
where $ \mathcal{B}_{i} $'s denote the Bell states associated with the Pauli matrices $ \sigma_{i} $'s by the following relation
\begin{equation}\label{a2}
	\mathcal{B}_{i}=\left(\sigma_{0}\otimes\sigma_{i}\right)\mathcal{B}_{0}\left(\sigma_{0}\otimes\sigma_{i}\right); \;i=1,2,3, 
\end{equation}
in which $ \sigma_{0}=I $, $ \sigma_{1}=\sigma_{x} $, $ \sigma_{2}=\sigma_{y} $, $ \sigma_{3}=\sigma_{z} $, and $I$ is the $2\times 2$ identity matrix.
For any arbitrary two-qubit system, each described by basis $\left\lbrace \Ket{0}, \Ket{1}\right\rbrace   $,  we have $ \mathcal{B}_{0}=\frac{1}{2}\left(\Ket{00} +\Ket{11}\right) \left(\Bra{00} +\Bra{11}\right)$, without loss of the generality.

\par
 Generalizing the above scenario, Lee and Kim \cite{lee2000entanglement} investigated the standard teleportation of  an unknown entangled state. In this protocol, the output state of the  teleportation channel is found to be
\begin{equation}\label{a3}
	\rho_{\text{out}}=\sum_{ij}p_{ij}\left( \sigma_{i}\otimes\sigma_{j}\right) \rho_{\text{in}}\left( \sigma_{i}\otimes\sigma_{j}\right),\  i,j=0,x,y,z,
\end{equation}
where $ p_{ij}=\text{Tr}\left[\mathcal{B}_{i}\rho_{\text{res}} \right]\text{Tr}\left[\mathcal{B}_{j}\rho_{\text{res}} \right] $ and $ \sum p_{ij}=1 $.

\subsubsection{Quantum estimation theory\label{QFI}}
First we  briefly review the principles of classical
estimation theory and the tools that  it provides to compute  the  bounds
to precision of any quantum metrology process.
 In an estimation problem we intend to   infer the value of a
parameter $ \lambda $ by measuring a related quantity $ X $. In order to 
solve the  problem, one should find an estimator $ \hat{\lambda}\equiv \hat{\lambda} (x_{1},x_{2},...)$, i.e., a real function of the measurement outcomes $ \left\{x_{k} \right\} $ to  the space of the possible values of the parameter $ \lambda $. In the classical scenario, the variance  $ \text{Var}(\lambda)=E[\hat{\lambda}^{2}] -E[\hat{\lambda}]^{2}$
of any unbiased estimator, in which  $ E[...] $ represents the mean with respect to the $ n $ identically
distributed random variables $ x_{i} $,  satisfies the
Cramer-Rao inequality $ \text{Var}(\lambda) \geq \dfrac{1}{MF_{\lambda}} $. It provides  a lower bound on the variance in terms of the
number of independent measurements $ M  $ and the Fisher information (FI) $ F(\lambda) $
\begin{equation}\label{cfi}
	F_{\lambda}=\sum_{x}\dfrac{[\partial_{\lambda}p(x|\lambda)]^{2}}{p(x|\lambda)},
\end{equation}
where $ p(x|\lambda) $ denotes the conditional probability of achieving the
value $ x $ when the parameter has the value $ \lambda $.  Here it is
assumed that the eigenvalue spectrum of observable $ X $ is discrete. If it is
continuous, the summation in Eq. (\ref{cfi}) must be replaced by
an integral.

\par In the quantum scenario $ p(x|\lambda)=\text{Tr}\left[ \rho P_{x}\right] $, where  $ \rho $ represents the state of the quantum system and $ P_{x} $ denotes the probability operator-valued measure (POVM) describing the measurement.  In summary,  it is possible to extract the value of the
physical parameter, intending to estimate it,  by measuring an observable $ X $ and then 
performing statistical analysis on the measurements results.
An efficient estimator  at least asymptotically saturates  the
Cramer-Rao bound.

Clearly, various observables give rise  to  miscellaneous
probability distributions, leading to different FIs and hence
to different precisions for estimating $ \lambda $. The quantum Fisher information (QFI), the ultimate
bound to the precision, is achieved by maximizing the FI over
the set of the observables. It is known that the set of projectors over the eigenstates of the SLD
 forms
an optimal POVM. 
The QFI of an unknown parameter $ \lambda $ encoded into the quantum state $ \rho\left(\lambda \right) $ is given by \cite{helstrom1969quantum,braunstein1996generalized}
\begin{equation}\label{01}
	\mathcal{F}_{\lambda}=\text{Tr}\left[\rho\left(\lambda \right)L^{2} \right]=\text{Tr}\left[\left( \partial_{\lambda}\rho\left(\lambda \right)\right) L\right], 
\end{equation}
in which $ L $ denotes the symmetric logarithmic derivative (SLD) given by $ \partial_{\lambda}\rho\left(\lambda \right)=\frac{1}{2}\left(L\rho\left(\lambda \right)+\rho\left(\lambda \right)L\right), $ where $ \partial_{\lambda}=\partial/\partial\lambda $.

Following the method presented in \cite{jing2014quantum,liu2016quantum} to calculate the QFI of block diagonal states $ \rho=\bigoplus^{n}_{i=1}\rho_{i}$, where $ \bigoplus $ denotes the direct sum, one finds that the SLD operator is also block diagonal and can be computed explicitly as $L=\bigoplus^{n}_{i=1}L_{i}$ in which $L_{i}$ represents the corresponding SLD operator for $\rho_{i}$. 
For 2-dimensional blocks, it is proved that the SLD operator for the $i$th block is expressed  by~\cite{liu2016quantum} 
\begin{equation}\label{Ldiagonal}
	L_{i}=\frac{1}{\mu_{i}}\left[\partial_{x}\rho_{i}+\xi_{i}\rho^{-1}_{i} -\partial_{x}\mu_{i} \right]. 
\end{equation}
in which $\xi_{i}=2\mu_{i}\partial_{x}\mu_{i}-\partial_{x}P_{i}/4$ where $\mu_{i}=\text{Tr}\left[\rho_{i}/2 \right]$ and $P_{i}=\text{Tr}\left[\rho^{2}_{i}\right]$. When  $\text{det} (\rho_{i})=0 $, $\xi_{i}$ vanishes.

\subsubsection{Quantum resources \label{QR}}
In this subsection we introduce some important measures to quantify key  resources which may be needed for implementing quantum information tasks.

\subsubsection*{Quantum entanglement \label{QE}}
 
 In bipartite qubit systems, the \textit{concurrence}  \cite{wootters1998entanglement,bennett1996mixed} is one of the most important measures used to quantify entanglement. 
 Introducing the "spin flip" transformation given by
 \begin{equation}
 	\rho\rightarrow \tilde{\rho}=(\sigma_{y}\otimes \sigma_{y} )\rho^{*}(\sigma_{y}\otimes \sigma_{y}),
 \end{equation}
 where $ * $ denotes the complex conjugate, Wootters \cite{wootters1998entanglement} presented
 the following analytic expression for the concurrence of the  mixed states of two-qubit systems:
 \begin{equation}
 C(\rho)= \text{max}(0,\lambda_{1}-\lambda_{2}-\lambda_{3}-\lambda_{4}),
 \end{equation}
  where $ \lambda_{i} $ are the square roots of the eigenvalues
 of $ \rho \tilde{\rho} $ in decreasing order. The concurrence has the range from from 0 to 1. 
 A quantum state with $  C = 0 $ is separable. Moreover, when $  C = 1 $, the state is maximally entangled.
For a X state the concurrence has the form
\begin{equation}\label{a8}
	C(\rho)=2\text{max}\left\lbrace 0,C_{1}(\rho),C_{2}(\rho) \right\rbrace ,
\end{equation}
where $ C_{1}(\rho)=|\rho_{14}|-\sqrt{\rho_{22}\rho_{33}}, C_{2}(\rho)=|\rho_{23}|-\sqrt{\rho_{11}\rho_{44}} $, and $ \rho_{ij} $'s are the elements of density matrix.

	\subsubsection*{Quantum coherence \label{QC}}

Quantum coherence (QC),   naturally a basis dependent concept  identified by the presence of off-diagonal terms in the density matrix,  is an important resource in quantum information theory (see\cite{streltsov2017colloquium} for a review). 
The reference basis
can be determined by the physics of the problem under investigation  or by the task for which the QC is required as a resource.
Although different measures  such as trace norm distance coherence \cite{shao2015fidelity}, $ l_{1} $ norm, and relative entropy of coherence \cite{baumgratz2014quantifying} are presented to quantify the QC of a quantum state, we adopt the $ l_{1} $ norm which is easily computable. For a quantum state with the density matrix $ \rho $, the $ l_{1} $ norm measure of quantum coherence \cite{baumgratz2014quantifying} quantifying the QC through the off diagonal elements of the density matrix in the reference basis, is given by

\begin{equation}\label{a7}
	\mathcal{C}_{l_{1}}\left(\rho \right)=\sum_{i,j\atop i\ne j}|\rho_{ij}|.
\end{equation}
In Ref. \cite{yao2015quantum} the authors investigated
whether a basis-independent measure of QC
can be defined or not. They found that the basis-free coherence is
equivalent to quantum discord \cite{ollivier2001quantum}, verifying the fact that
coherence can be introduced as a form of quantum correlation in multi-partite
quantum systems.

\subsubsection*{Quantum discord}
Quantum discord (QD) \cite{ollivier2001quantum}, defined as difference
between total correlations and classical correlations \cite{fanchini2017lectures}, represents the quantumness of the state of a quantum system. QD is a resource for certain quantum technologies \cite{pirandola2014quantum}, because it can be preserved for a long time even when entanglement exhibits a sudden death.
 Usually, computing QD for a general state is not easy because it involves the optimization of the classical correlations. Nevertheless, for a two-qubit $ X $ state system, an easily computable  expression of QD is given by \cite{wang2010classical}

\begin{equation}\label{a9}
	QD(\rho)=\text{min}\left(Q_{1},Q_{2} \right),
\end{equation}
in which
\begin{eqnarray}\label{a10}
	\nonumber Q_{j}=H\left(\rho_{11} +\rho_{33}  \right)+\sum_{i=1}^{4}\lambda_{i}\text{log}_{2}\lambda_{i}+D_{j},~~~\left(j=1,2\right),~~~~~\\
	D_{1}=H\left(\frac{1+\sqrt{\left[1-2\left( \rho_{33} +\rho_{44}\right)  \right] ^{2}+4\left(|\rho_{14}|+|\rho_{23}| \right) ^{2}}}{2} \right), \\\nonumber
	D_{2}=-\sum_{i}\rho_{ii}\text{log}_{2}\rho_{ii}-H\left(\rho_{11} +\rho_{33}  \right),~~~~~~~~~~~~~~~~~~~~~~~~~~\\\nonumber
	\nonumber H\left(x \right)=-x\text{log}_{2}x-\left(1-x \right)\text{log}_{2}\left(1-x \right),~~~~~~~~~~~~~~~~~~~~~~~~  
\end{eqnarray}
and $ \lambda_{i} $'s represent the eigenvalues of the bipartite density matrix $ \rho $.

\subsubsection{Hilbert-Schmidt speed}\label{HILBERT-SCHMIDT SPEED}

First we consider  the distance measure $ d(p,q) $, defined as \cite{gessner2018statistical}   
	$ [d(p,q)]^{2}=1/2\sum\limits_{x}^{}|p_{x}-q_{x}|^{2}, $
where $ p = \{p_{x}\}_{x} $ and $ q = \{q_{x}\}_{x} $, depending on parameter $ \varphi $, represent the probability distributions, leading to the classical statistical speed
$
	s[p(\varphi_{0})]=\dfrac{\mathrm{d}}{\mathrm{d}\varphi}\ d(p(\varphi_{0}+\varphi),p(\varphi_{0})).
$
In order to extend these classical notions to the quantum case, 
one can consider  a given pair of quantum states $ \rho $ and $ \sigma $, and  write $ p_{x} = \text{Tr}\left[E_{x}\rho\right] $ and $ q_{x} = \text{Tr}\left[E_{x}\sigma\right] $ which denote the measurement
probabilities corresponding to the positive-operator-valued measure (POVM) $ \{E_{x}\geq 0\} $ satisfying $\sum\limits_{x}^{} E_{x} = \mathbb{I}  $ .
Maximizing the classical distance $ d(p,q) $ over all possible choices of POVMs \cite{PhysRevA.69.032106}, we can obtain the corresponding quantum distance called  the Hilbert-Schmidt  distance $ \delta_{HS} $ \cite{ozawa2000entanglement} given by
$
	\delta_{HS} (\rho,\sigma)\equiv \max_{\{E_{x}\}}d(p,q)=\sqrt{\frac{1}{2}\text{Tr}[(\rho-\sigma)^{2}]}.
$
Therefore, the Hilbert-Schmidt speed (HSS), the corresponding quantum statistical speed, is achieved by maximizing the classical statistical speed  over all possible POVMs \cite{paris2009quantum,gessner2018statistical}
\begin{align}\label{HSSS}
	HS\!S \big(\rho(\varphi)\big)\equiv HS\!S_{\varphi}
	&=\max_{\{E_{x}\}} \text{s}\big[p(\varphi)\big]\nonumber\\
	&=\sqrt{\frac{1}{2}\text{Tr}\bigg[\bigg(\dfrac{\mathrm{d}\rho(\varphi)}{\mathrm{d}\varphi}\bigg)^2\bigg]}.
\end{align}
 \section{Dynamics of the non-localized qubit realized by Majorana modes   \label{Model}}

  We discuss the time evolution of a topological qubit realized by spatially-separated Majorana modes,  and placed on top of an s-wave superconductor. The Majorana modes are generated at the endpoints of some nanowire with strong spin-orbit
  interaction, and are  subject to an external magnetic field \textbf{$\mathcal{B}$}  along the wire axis direction.
 They   are independently 
coupled to  metallic nanowires via tunnel junctions  such that the  tunneling strengths  are controllable through  external gate voltages. 
\par
The total Hamiltonian can be written as \cite{ho2014decoherence}
\begin{equation}
	H=H_{S}+H_{E}+V 
\end{equation}
in which $ H_{s} $ represents the Hamiltonian of the topological qubit. Assuming  that the Majorana modes  are zero-energy ones, we can put $ H_{S}=0 $.
 Furthermore,  the Hamiltonian of the  environment,  composed of 1D electrons,  is denoted by $ H_{E} $ which can be written in terms of  the electrons's creation  and annihilation operators, i.e.,  $\Xi^{\dagger}_{j} $, and $\Xi_{j} $, respectively. The noise affecting  the topological qubit  can be modelled as a fermionic
Ohmic-like environment realized by placing a metallic
nanowire close to the Majorana endpoint and  described by spectral density $ \rho_{spec}\propto \omega^{Q} $ with $ Q\geq 0 $.   The environment is called Ohmic for $ Q = 1 $, and super (sub)-Ohmic for $ Q > 1~(Q < 1)$.  A physical implementation of  this kind of 
environment  is the helical Luttinger liquids realized
as interacting\textit{ edge states} of two-dimensional topological insulators \cite{chao2013nonequilibrium}. 
Moreover, ignoring the weak coupling between the Majorana modes and the  higher order terms involving greater numbers of
Majorana modes, the system-environment interaction Hamiltonian is given by
\begin{equation}
	V\simeq B(\gamma_{1}\mathcal{O}_{1}+\gamma_{2}\mathcal{O}_{2}),
\end{equation} 
  in which $ B $ is the real coupling strength  between the Majorana modes and the environment,  and $ \mathcal{O}_{1(2)} $
   denotes the composite
  operator of  $\Xi^{\dagger}_{j} $, and $\Xi_{j} $. In addition,
 the localized Majorana modes $ \gamma_{1} $  
and $ \gamma_{2} $   satisfy the following properties:
\begin{equation}\label{MqubitProp}
	\gamma^{\dagger}_{a}=\gamma_{a},~~~\{\gamma_{a},\gamma_{b}\}=2\delta_{ab},
\end{equation}
where $ a,b=1,2 $.  $ \gamma_{1,2} $ can be represented by:
\begin{equation}\label{MajorPauli}
	\gamma_{1}=\sigma_{1},~~~\gamma_{2}=\sigma_{2},~~~\text{i}\gamma_{1}\gamma_{2}=\sigma_{3},
\end{equation}
wherein $ \sigma_{j} $'s denote the Pauli matrices.
Besides, the two
Majorana modes can be treated as a  topological (non-local) qubit described by basis states $ |0\rangle $ and $ |1\rangle $ such that:
\begin{equation}\label{Majorqubit}
	\frac{1}{2}(\gamma_{1}-\text{i}\gamma_{2})|0\rangle=|1\rangle,~~~~~\frac{1}{2}(\gamma_{1}+\text{i}\gamma_{2})|1\rangle=|0\rangle.
\end{equation}

\par
The fermionic Ohmic-like environment environment, composed of 1D electrons, is either Fermi or Luttinger liquid whose interaction strengths are characterized by the parameter $ \kappa=(Q+1)/2 $.   Therefore, the larger $ \kappa $, the stronger the correlation/interaction exhibited by the Luttinger liquid nanowire. It should be noted that $ \kappa \equiv (K+1/K)/4 $ in which $ K $, representing the   \textit{Luttinger parameter}, can be roughly  estimated as
$ K \thicksim (1+\dfrac{U}{2\epsilon_{F}})$
in which  $\epsilon_{F} $ is the Fermi energy and $U$ denotes the characteristic Coulomb energy of the wire \cite{kane1992transport,ho2014decoherence}. Accordingly,   the value
of $ \kappa $ and consequently   $ Q $  can be tuned by changing the effective attractive/repulsive short range interactions in the
wire.

\par
 Assuming that the the Majorana qubit is initially prepared in the following state:

\begin{equation}\label{InitialMajorqubit}
	\varrho(0)=\left(\begin{array}{cc}
		\varrho_{11}(0)&\varrho_{12}(0)  \\
		\varrho_{21}(0)&\varrho_{22}(0)  \\
	\end{array}\right),
\end{equation}
one can find that the reduced
density matrix at time $ t $ is given by (for details, see \cite{ho2014decoherence}):

\begin{equation}\label{Reduced1qMajorqubit}
	\varrho(t)=\frac{1}{2}\left(\begin{array}{cc}
		1+(2\varrho_{11}(0)-1)\alpha^{2}(t)&2\varrho_{12}(0)\alpha(t)  \\
		2\varrho_{21}(0)\alpha(t)&1+(2\varrho_{22}(0)-1)\alpha^{2}(t) \\
	\end{array}\right),
\end{equation}
in which

\begin{equation}\label{alpha}
	\alpha(t)=\text{e}^{-2B^{2}|\beta|I_{Q}},~~~~~\beta\equiv \dfrac{-4\pi}{\Gamma(Q+1)}(\dfrac{1}{\varGamma_{0}})^{Q+1},
\end{equation}
where $ \varGamma_{0} $ denotes   the cutoff frequency, appeared in the  \textit{Green function} for the fermionic environment, and $ \Gamma (z) $ represents the Gamma function. Furthermore, 
\begin{equation}\label{IQ}
	I_{Q}(t)= \left\{
	\begin{array}{rl}
		2\varGamma_{0}^{Q-1} \Gamma(\frac{Q-1}{2})\bigg[1-\,_1F_1\big(\frac{Q-1}{2};\frac{1}{2};-\frac{t^{2}\varGamma^{2}_{0}}{4}\big)\bigg]&~~~~~~~~ \text{for}~ Q\neq 1,\\
		\frac{1}{2}t^{2}\varGamma^{2}_{0} \,_2F_2\bigg(\{1,1\};\{3/2,2\};-\frac{t^{2}\varGamma^{2}_{0}}{4}\bigg)   & ~~~~~~~~\text{for}~ Q=1,
	\end{array} \right.
\end{equation}
where $ \,_pF_q $ represents the \textit{generalized hypergeometric function}.
It should be  noted  that
$ \varrho^{T}(0) $, describing the initial state of the total system, is  assumed to  be uncorrelated, i.e.,  $\varrho^{T}(0)=\varrho(0)\otimes \varrho_{E}$,
where $\rho_{S}(0)$ and $\rho_{E}  $ represent, respectively, the initial density matrix of the the topological qubit and that of the environment.

 The trace norm defined by $\parallel \rho \parallel=\text{Tr}\sqrt{\rho^{\dagger}\rho} =\sum\limits_{k}\sqrt{\lambda_{k}} $, where $ \lambda_{k} $'s represent the eigenvalues of $\rho^{\dagger}\rho  $ can be used to define the \textit{trace distance} \cite{fanchini2017lectures} $ D(\rho^{1},\rho^{2})=\frac{1}{2}\parallel \rho^{1}-\rho^{2} \parallel $  an important measure of the  distinguishability between two  quantum states $ \rho^{1} $ and $ \rho^{2} $. 
This measure was applied by Breuer, Laine, and
Piilo (BLP) \cite{breuer2009measure} to define one of the most  important characterization  of non-Markovianity in quantum systems. They   proposed  that a non-Markovian
process can be characterized by a backflow of information from the
environment  into the open system  mathematically detected by the witness $  \dot{D}\big(\rho^{1}(t),\rho^{2}(t)\big)>0 $ in which $ \rho^{i}(t) $ denotes the evolved state starting from the initial state  $ \rho^{i}(0) $, and the dot represents the time derivative.

 \par
In order  to calculate the BLP measure  of non-Markovianity, one has to find
a specific pair of optimal initial states maximizing the time
derivative of the trace distance. For
any non-Markovian evolution of a one-qubit system, it is known that the maximal
backflow of information occurs for a pair of pure orthogonal
initial states corresponding to antipodal points on the surface
of the Bloch sphere of the qubit. 
In Ref. \cite{jahromi2019quantum},  the non-Markovian features of our model has been investigated and found that
  the optimal initial states are given by $ \{  |0\rangle, |1\rangle \}$, and   the corresponding evolved trace
distance is obtained as  $ D\big(\rho^{1}(t),\rho^{2}(t)\big)\equiv D(t)=\alpha^{2}(t) $.  
\par
Here we show that those features  can also be extracted   by another faithful witness of non-Markovianity recently proposed in \cite{jahromi2020witnessing}. According to this witness, at first we should compute the evolved density matrix  $\varrho(t)$ corresponding to   initial state
 $
	|\psi_{0}\rangle=\frac{1}{2}\bigg(\text{e}^{i\varphi} \ket{+}+\ket{-} \bigg)
$. Here  $ \{\ket{+},\ket{-}\} $, constructing a complete and orthonormal set (basis) for the qubit, is usually associated with the computational basis. Then the positive changing
rate of the Hilbert-Schmidt speed (HSS), i.e.,   $ \dfrac{\text{d}HS\!S_{\varphi}(\varrho(t))}{\text{d}t}>0 $ in which the HSS has been computed with respect to the initial phase $ \varphi $, can  identify the memory effects. This witness of non-Markovianity is in total
agreement with the  BLP witness,  thus
detecting the system-environment information backflows. 

\par
Considering the initial state 
$
|\psi_{0}\rangle=\frac{1}{2}\bigg(\text{e}^{i\varphi} \ket{0}+\ket{1} \bigg)
$ in our model, and using Eq. (\ref{HSSS}), we find that the HSS of $ \varrho(t) $, obtained from Eq. (\ref{Reduced1qMajorqubit}), with respect to the phase parameter $ \varphi $
is given by

\begin{equation}\label{HSS1}
HS\!S_{\varphi}(\varrho(t))\equiv	HS\!S_{\varphi}(t)=\dfrac{\alpha(t)}{2},
\end{equation}
leading to  
\begin{equation}\label{QFIHSSQ2}
	HS\!S_{\varphi}(t)=\dfrac{\sqrt{D(t)}}{2}
	\Rightarrow 
	\dfrac{\text{d}HS\!S_{\varphi}(t)}{\text{d}t} = \frac {1}{4\sqrt{D(t)}} \dfrac{\text{d}D(t)}{\text{d}t}. 
\end{equation}
Because the trace distance is a nonnegative quantity, the  signs of $ \dfrac{\text{d}D(t)}{\text{d}t} $ and $ \dfrac{\text{d}HS\!S_{\varphi}}{\text{d}t} $ coincide and hence they exhibit the same 
qualitative dynamics. 
\par
This result verifies the fact that 
the HSS-based measure is in perfect agreement with the trace
distance-based witness and can be used as an efficient as well as easily computable tool for
detecting the non-Markovianity in our model. 

\par 
A necessary condition which should be satisfied by a faithful witness of non-Markovianity is contractivity under Markovian dynamics.
It should be noted that the noncontractivity of the Hilbert-Schmidt distance does not consequently lead to noncontractivity of the HSS. In more detail, as explained in Sec. \ref{HILBERT-SCHMIDT SPEED},    the Hilbert-Schmidt distance is calculated by maximizing over all the possible choices of POVMs $\{E_x\}$ of the adopted distance measure, while the HSS is determined by maximization applied after the differentiation with respect to $\varphi$, starting from the corresponding distance measure. Because of these computational subtleties, from  noncontractivity of the Hilbert-Schmidt distance we cannot deduce that   the HSS is also noncontractive.
Moreover, in Ref. \cite{jahromi2020witnessing} we have demonstrated that the HSS is contractive in Hermitian systems.
\section{teleportation using non-localized qubits }\label{Tel}
In this section, we investigate the scenario in which the  non-localized qubits are used as the resource for teleportation of an entangled pair. 
\subsection{Output state of the teleportation}\label{Tel1}
Calculating the  the eigenvalues and eigenvectors of the \textit{Choi matrix} \cite{leung2003choi} of the map 
$ \Phi_{t} $ satisfying the relationship $\Phi_{t} (\rho(0))=\rho(t)  $, one can obtain  the corresponding operator-sum representation $\varrho(t)=\sum_{i=1}^4 K_i(t) \varrho(0) K_i(t)^\dagger$ where the Kraus operators $\{K_{i}(t)\}  $ are given by \cite{jahromi2021hilbert}
\begin{align}\label{Kraus opt}
	&K_{1}(t)=\left(
	\begin{array}{cc}
		\frac{\alpha -1}{2} & 0 \\
		0 & \frac{1-\alpha }{2} \\
	\end{array}
	\right),~K_{2}(t)=\left(
	\begin{array}{cc}
		\frac{\alpha +1}{2} & 0 \\
		0 & \frac{\alpha +1}{2} \\
	\end{array}
	\right),\nonumber\\
	&K_{3}(t)=\left(
	\begin{array}{cc}
		0 & \frac{\sqrt{1-\alpha ^2}}{\sqrt{2}} \\
		0 & 0 \\
	\end{array}
	\right),~K_{4}(t)=\left(
	\begin{array}{cc}
		0 & 0 \\
		\frac{\sqrt{1-\alpha ^2}}{\sqrt{2}} & 0 \\
	\end{array}
	\right).
\end{align}
In the following, we take a system formed by two noninteracting non-localized qubits  such that each of which locally interacts with its environment in Sec. \ref{Model}. 
\begin{figure} [t!]
	\centering
	\includegraphics[width=0.5\textwidth]{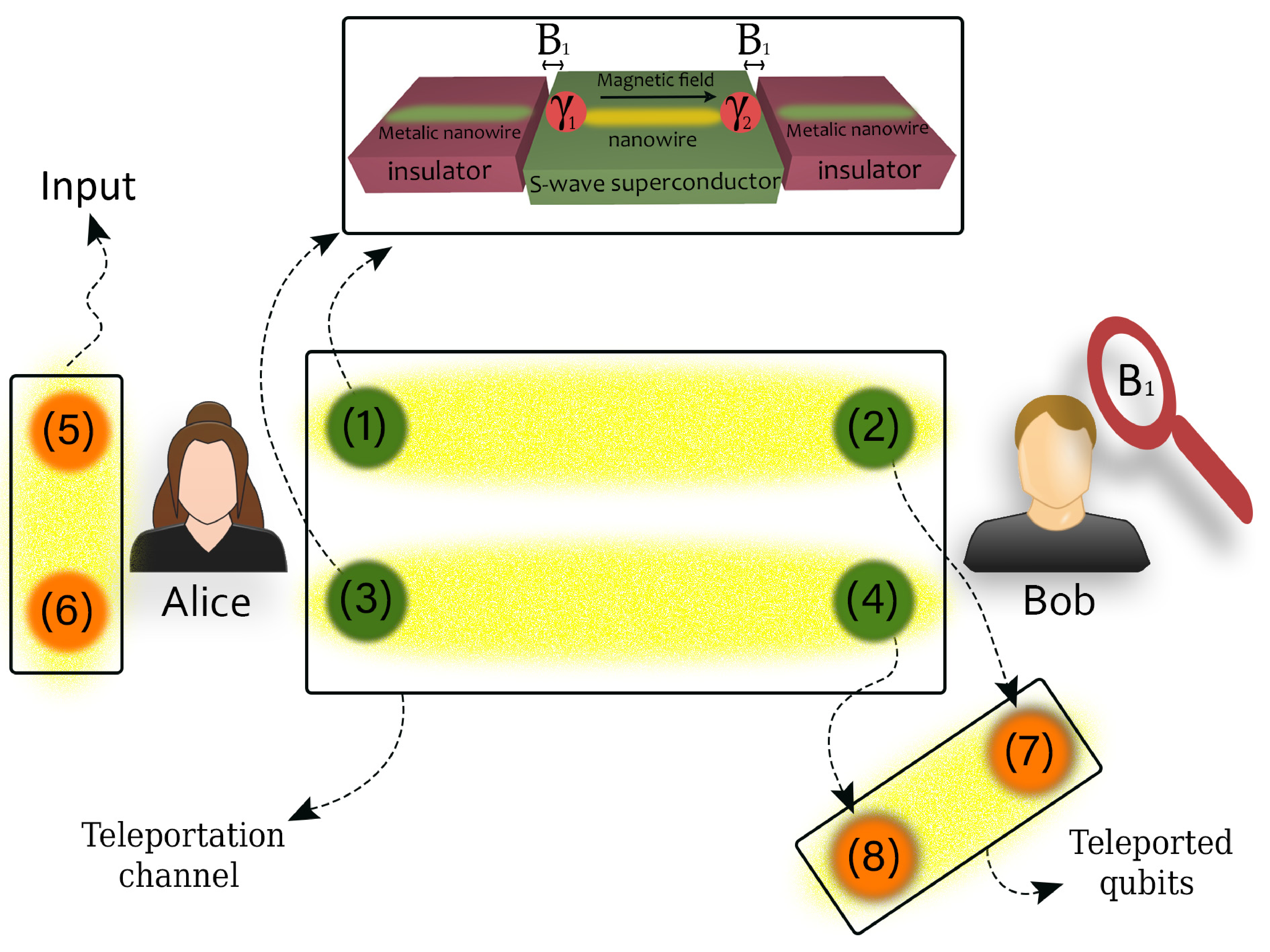}
	\caption{\textbf{Sketch of the two-qubit noisy  teleportation protocol through non-localized Majorana qubits.} Qubits (1,2) and (3,4), representing two copies of the evolved noisy two-qubit system, are used as the resource channel for the protocol, and  shared between Alice and Bob. The input qubits (5,6) which should be  teleported,  are manipulated in Alice's location. After the protocol, Bob's qubits, coming from the resource state, are transformed into qubits (7,8) representing the teleported qubits. Bob  employs these qubits for remote sensing to estimate $ B_{1} $, i.e., the strength  of coupling  between Alice's Majorana qubit and its fermionic environment.}\label{fig:2qubittel}
\end{figure}
  Because the environments are independent in our model, the Kraus operators, describing the dynamics of this composite system, are just the tensor products of Kraus operators acting on each of the qubits \cite{nielsen2002quantum}. Therefore, assuming that the two-qubit system is  prepared in  the initial  entangled state $ \rho_{0}=|\Psi_{0}\rangle \langle \Psi_{0}| $ where 
  \begin{equation}\label{initialtwoqubit}
  	|\Psi_{0}\rangle=\cos(\vartheta/2)|00\rangle+\sin(\vartheta/2)|11\rangle.
  \end{equation} 
  we can find that the nonzero  elements of the
  evolved density matrix  $ \rho(t) $ are given by
  \begin{eqnarray}
  	&&\rho_{1,1}(t)=\frac{1}{4} \left(\left(\text{$\alpha $}_{1}^2+\text{$\alpha $}_{2}^2\right) \cos (\vartheta )+\text{$\alpha $}_{1}^2 \text{$\alpha $}_{2}^2+1\right),\nonumber \\
  	&& \rho_{22}(t)=\frac{1}{4} \left(\left(\text{$\alpha $}_{1}^2-\text{$\alpha $}_{2}^2\right) \cos (\vartheta )-\text{$\alpha $}_{1}^2 \text{$\alpha $}_{2}^2+1\right),\\
  	&& \rho_{33}(t)=\frac{1}{4} \left(\left(\text{$\alpha $}_{2}^2-\text{$\alpha $}_{1}^2\right) \cos (\vartheta )-\text{$\alpha $}_{1}^2 \text{$\alpha $}_{2}^2+1\right),\nonumber\\
  	&&\rho_{4,4}(t)=1-\bigg(\rho_{1,1}(t)+\rho_{2,2}(t)+\rho_{3,3}(t)\bigg),\nonumber\\
  	&&\rho_{1,4}(t)=\frac{1}{2} \text{$\alpha $}_{1} \text{$\alpha $}_{2} \sin (\vartheta ),\nonumber
  \end{eqnarray}
where $ \alpha_{i} $ ($ i=1,2 $), presented in Eq. (\ref{alpha}) for each qubit, includes the Ohmicity parameter  $ Q_{i} $, cutoff $ \Gamma_{i} $, and coupling (tunneling)  strength $ B_{i} $.
\par
Sharing two copies of this two-qubit system between Alice and Bob, as the resource or channel ($ \rho(t)=\rho_{ch} $)  for the quantum teleportation of the input state
 \begin{equation}\label{inputstate}
  |\psi_{in}\rangle=\text{cos}~ \left(\frac{\theta}{2} \right)  |10\rangle+e^{i\phi}~\text{sin}~\left(\frac{\theta}{2} \right)  |01\rangle, 
    \end{equation} 
  with $\ 0\leq\theta\leq\pi,\ 0\leq\varphi\leq2\pi$, and  following Kim and Lee's two-qubit teleportation protocol \cite{PhysRevLett.84.4236},  we obtain the output state  as
  \begin{widetext}
  \begin{equation}\label{outputstate}
  \rho_{\text{out}}(t)=	\left(
  \begin{array}{cccc}
  	\frac{1}{4} \left(1-\alpha ^4\right) & 0 & 0 & 0 \\
  	0 & \frac{1}{4} \left(-2 \alpha ^2   \cos \theta+\alpha ^4+1\right) & \frac{1}{2} \alpha ^2 e^{i \phi }   \sin \theta    \sin ^2\vartheta & 0 \\
  	0 & \frac{1}{2} \alpha ^2 e^{-i \phi }   \sin \theta   \sin ^2 \vartheta & \frac{1}{4} \left(2 \alpha ^2   \cos \theta +\alpha ^4+1\right) & 0 \\
  	0 & 0 & 0 & \frac{1}{4} \left(1-\alpha ^4\right)
  \end{array}
  \right),
  \end{equation} 
       \end{widetext}
  where  $\alpha(t) =\alpha_{1}(t)\alpha_{2}(t) $.
 Extracting the results throughout this paper, we assume that the two-qubit states (\ref{initialtwoqubit}) and (\ref{inputstate}) are maximally entangled, i.e., $ \theta=\vartheta=\pi/2 $.

\subsection{Remote sensing through teleported qubits}\label{Magneticsensing}

We address the estimation of the coupling strength $ B_{1} $  at Alice's location, through the non-localized qubits, teleported to Bob's location. It is an interesting model motivating investigation  of making remote  sensors (see Fig. \ref{fig:2qubittel}). 

\par The state of the two-qubit system, used for the remote sensing, is given by Eq. (\ref{outputstate}). It can be transformed into a block-diagonal form by  changing the order of the basis vectors. Then, using Eq. (\ref{Ldiagonal}), we find that the SLD, associated with  the coupling strength, is obtained as
 \begin{widetext}
\begin{equation}\label{LSLD}
L(t)=	\left(
	\begin{array}{cccc}
		\frac{4 \alpha ^3}{\alpha ^4-1} & 0 & 0 & 0 \\
		0 & \frac{4 \left(-2 \alpha ^3 \left(\sin^2 \theta   \right) \left(\sin ^4 \vartheta  \right)+\alpha  (\cos \theta) \left(-2 \alpha ^2 (\cos \theta   )+\alpha ^4-1\right)+\alpha ^7+\alpha ^3\right)}{\left(\alpha ^4+1\right)^2-4 \alpha ^4 \left(\left(\sin ^2 \theta  \right) \left(\vartheta  \sin ^4\right)+\cos ^2 \theta  \right)} & -\frac{4 \alpha  \left(\alpha ^4-1\right) e^{i \phi } (\sin \theta   ) \left(\vartheta  \sin ^2\right)}{\left(\alpha ^4+1\right)^2-4 \alpha ^4 \left(\left(\sin ^2 \theta  \right) \left(\vartheta  \sin ^4\right)+\cos ^2 \theta  \right)} & 0 \\
		0 & -\frac{4 \alpha  \left(\alpha ^4-1\right) e^{-i \phi } (\sin \theta   ) \left(\vartheta  \sin ^2\right)}{\left(\alpha ^4+1\right)^2-4 \alpha ^4 \left(\left(\sin ^2 \theta  \right) \left(\vartheta  \sin ^4\right)+\cos ^2 \theta  \right)} & \frac{4 \alpha  \left(-2 \alpha ^2 \left(\sin ^2 \theta  \right) \left(\vartheta  \sin ^4\right)-(\cos \theta   ) \left(2 \alpha ^2 (\cos \theta   )+\alpha ^4-1\right)+\alpha ^6+\alpha ^2\right)}{\left(\alpha ^4+1\right)^2-4 \alpha ^4 \left(\left(\sin ^2 \theta  \right) \left(\vartheta  \sin ^4\right)+\cos ^2 \theta  \right)} & 0 \\
		0 & 0 & 0 & \frac{4 \alpha ^3}{\alpha ^4-1} \\
	\end{array}
	\right)\dfrac{\partial \alpha(t)}{\partial B_{1}}.
\end{equation}
 \end{widetext}
Inserting the SLD into Eq. (\ref{01}) leads to the following expression for  the QFI with respect to $ B_{1} $:
\begin{equation}
	\mathcal{F}_{B_{1}}(t)=\frac{8 \alpha^2(t) }{1-\alpha ^4(t)}\left(\dfrac{\partial \alpha(t)}{\partial B_{1}}\right)^{2}.
\end{equation}
 The (normalized) eigenstates of $ L(t) $ are given by
\begin{eqnarray}
	&&\ket{\Psi_{1}}=\left(
	\begin{array}{c}
		0 \\
		\frac{e^{i \phi } \csc ^2 \vartheta ~ \left(\eta-2 \cot \theta   \right)}{\sqrt{\csc ^4 \vartheta ~  \left(\eta-2 \cot \theta   \right)^2+4}} \\
		\frac{2}{\sqrt{\csc ^4 \vartheta ~ \left(\eta-2 \cot\theta   \right)^2+4}} \\
		0 \\
	\end{array}
	\right),\nonumber  \ket{\Psi_{2}} =\left(
	\begin{array}{c}
		0 \\
		\frac{-e^{i \phi } \csc ^2 \vartheta~   \left(\eta+2 \cot \theta  \right)}{\sqrt{\csc ^4 \vartheta ~  \left(\eta+2 \cot \theta   \right)^2+4}} \\
		\frac{2}{\sqrt{\csc ^4 \vartheta ~  \left(\eta+2\cot  \theta   \right)^2+4}} \\
		0 \\
	\end{array}
	\right),\\
	&& \ket{\Psi_{3}}=\left(
	\begin{array}{c}
		0 \\
		0 \\
		0 \\
		1 \\
	\end{array}
	\right), \ket{\Psi_{4}}=\left(
	\begin{array}{c}
		1 \\
		0 \\
		0 \\
		0 \\
	\end{array}
	\right),
\end{eqnarray}
where $ \eta(\theta,\vartheta)=\sqrt{4 \csc ^2 \theta  +2 (\cos (2 \vartheta )  -3) \left(\cos ^2 \vartheta  \right)} $.
Because the set of the projectors over the above eigenstates, i.e., $\left\{P_{i}\equiv \ket{\Psi_{i}(t)}\bra{\Psi_{i}(t)}; ~i=1,2,3,4 \right\}    $, constructs
an optimal POVM,  the corresponding FI  saturates the QFI. In order to explicitly see  this interesting fact,  we first compute the conditional probabilities $p_{i}=\text{Tr}\left[\rho_{out}P_{i}\right]  $ appearing in Eq. (\ref{cfi}), leading to following expressions:
\begin{eqnarray}
	&&p_{1}=\dfrac{\left(\alpha ^2  \eta \sin (\theta) +\alpha ^4+1\right)}{4} 
	,\nonumber \\
	&& p_{2}=\dfrac{\left(-\alpha ^2 \eta\sin (\theta ) +\alpha ^4+1\right)}{4},\\
	&& p_{3}=\frac{1}{4} \left(1-\alpha ^4\right),\nonumber\\
	&&p_{4}=\frac{1}{4} \left(1-\alpha ^4\right).\nonumber
\end{eqnarray}
A straightforward calculation shows that 
\begin{equation}\label{cfi}
	F_{B_{1}}(t)=\sum_{i}\dfrac{[\partial_{B_{1}}p_{i}]^{2}}{p_{i}}=\mathcal{F}_{B_{1}}(t).
\end{equation}
 \begin{figure}[ht!]
	\subfigure[]{\includegraphics[width=9cm]{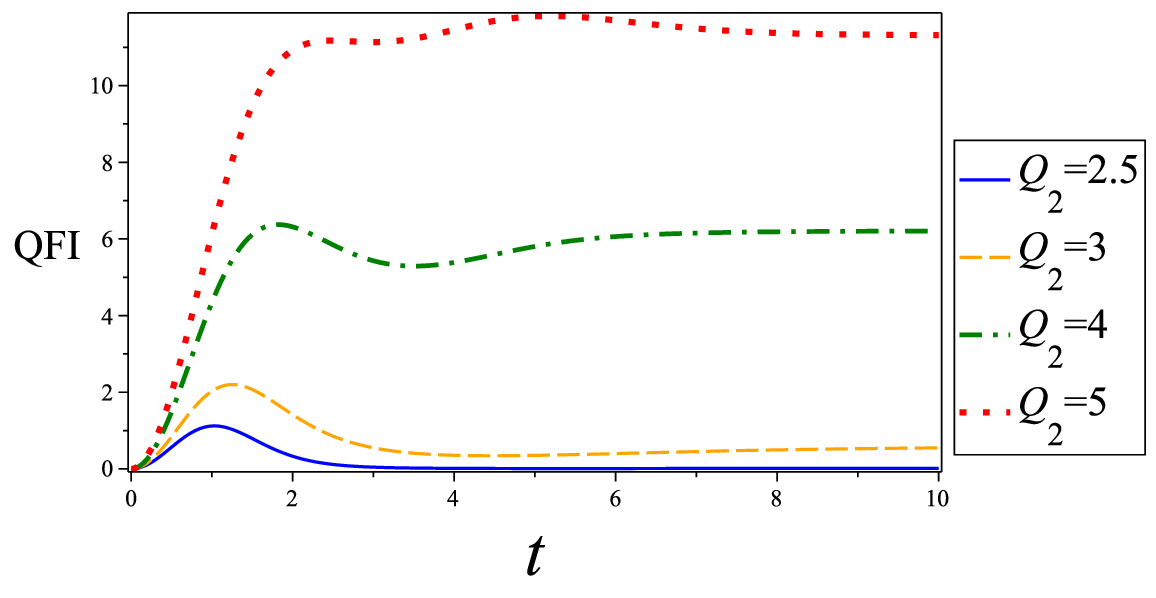}\label{Figg1a} }
	\subfigure[]{\includegraphics[width=9cm]{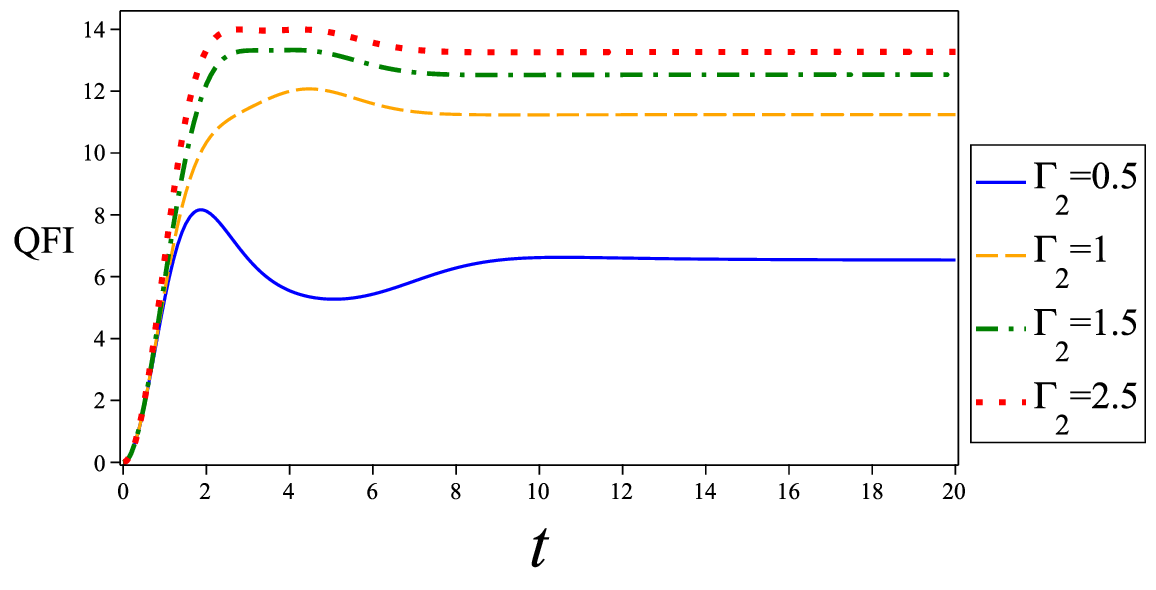}\label{Figg1b} }
	\caption{Dynamics of the quantum Fisher information  (QFI)  associated with $ B_{1} $ as a function of (a)   Ohmicity parameter $ Q_{2} $ and (b)  cutoff frequency  $\Gamma_{2}  $ controlled by Bob implementing the quantum sensing.}
	\label{fig1} 
\end{figure}
\par
Because the quantum metrology process is implemented by Bob, it is reasonable to assume that we can control  the environmental parameters  at Bob's location, while the same parameters at Alic's location are out of our control. Therefore, we investigate how the remote  sensing  can be enhanced by controlling the interaction of working qubits in the process of the measurement. 
\par

\par
The time variation of the QFI for different values of the Ohmicity parameter and cutoff frequency is plotted in Fig. \ref{fig1}. It shows that increasing the Ohmicity parameter $ Q $ enhances the quantum estimation of the coupling strength and frustrates  the QFI degradation (see Fig. \ref{Figg1a}). In fact, with increasing $ Q $,  after
a while, the QFI becomes approximately constant over time, a phenomenon known as
\textit{QFI trapping}.
Therefore, preparing a super-Ohmic environment at the location of the teleported qubits is the best strategy to extract information about the Alice's coupling strength. 
The super-Ohmic
environment frequently describes the effect of the interaction between a charged particle and its electromagnetic
field. It should be noted  that such engineering of
the ohmicity of the spectrum can be implemented experimentally \cite{haikka2014non}, e.g., when
simulating the dephasing model in trapped ultracold atoms, as described in Ref. \cite{haikka2011quantifying}.
\par
Figure \ref{Figg1b}  exhibits the important role of the cutoff frequency $ \Gamma_{2} $, imposed on the Bob's working qubits, in upgrading the sensors. Intensifying the cutoff frequency, Bob can   retard the QFI
loss  and therefore enhance the remote  sensing.  Generally, the cutoff frequency $ \Gamma_{2} $ is associated to a characteristic time $ \tau_{c}\sim \Gamma_{2}^{-1}  $
setting the fastest  time scale of the irreversible
dynamics. $ \tau_{c} $ is usually known as the correlation time of the environment \cite{benedetti2018quantum,salari2019quantum}.
In our model, Bob can achieve the QFI trapping with imposing a high cutoff frequency.

\par
Overall, in order to achieve the best efficiency in the process of remote  sensing, Bob should effectively control the Ohmicity parameter as well as the cutoff frequency. 
In Ref. \cite{jahromi2019quantum}, it has been discussed that how appropriate values of these parameters
allow a transition from Markovian to non-Markovian qubit dynamics.
Nevertheless, our results hold for both Markovian and non-Markovian evolution  of the the two-qubit systems used as the teleportation resources.

Our numerical computation shows that intensifying the coupling strength $ B_{2} $ causes  the QFI  to be suppressed,  reducing the optimal
precision of the estimation (see Fig. \ref{conB2}). Therefore, Bob should decrease the intensity of  coupling strength in order to provide a proper condition for the quantum  sensors. Tuning the coupling strengths is achievable in the experiments, by
controlling the gate voltages of the  tunneling junctions.
\par
In Sec. \ref{entanglement}, we investigate one of the physical origins of these phenomena.

      \subsection{ Feasible measurement for optimal estimation   \label{PI}}
Another important question is how we can physically design
the optimal estimation, i.e., a practically feasible measurement  whose Fisher information is equal to the QFI. Reminding that  the optimal POVM can be constructed by the eigenvectors of the SLD, we should compute them and check whether these eigenvectors coincide with those of  some observable of the system or  not.
Although in the general case this coincidence is not completely achievable, we interestingly find that with high cutoff frequency $ \Gamma_{1} $ or large values of $ Q_{1} $,   the optimal POVM can be \textit{approximately} constructed
by the eigenvectors of $ \sigma_{z} \otimes  \sigma_{z}$.
\textit{ It means that measurement of $ \sigma_{z} $ on each qubit leads to the near-optimal remote estimation, saturating the quantum upper bound.}

\subsection{Quantum entanglement between teleported sensors}\label{entanglement}
\begin{figure} [t!]
	\centering
	\includegraphics[width=0.47\textwidth]{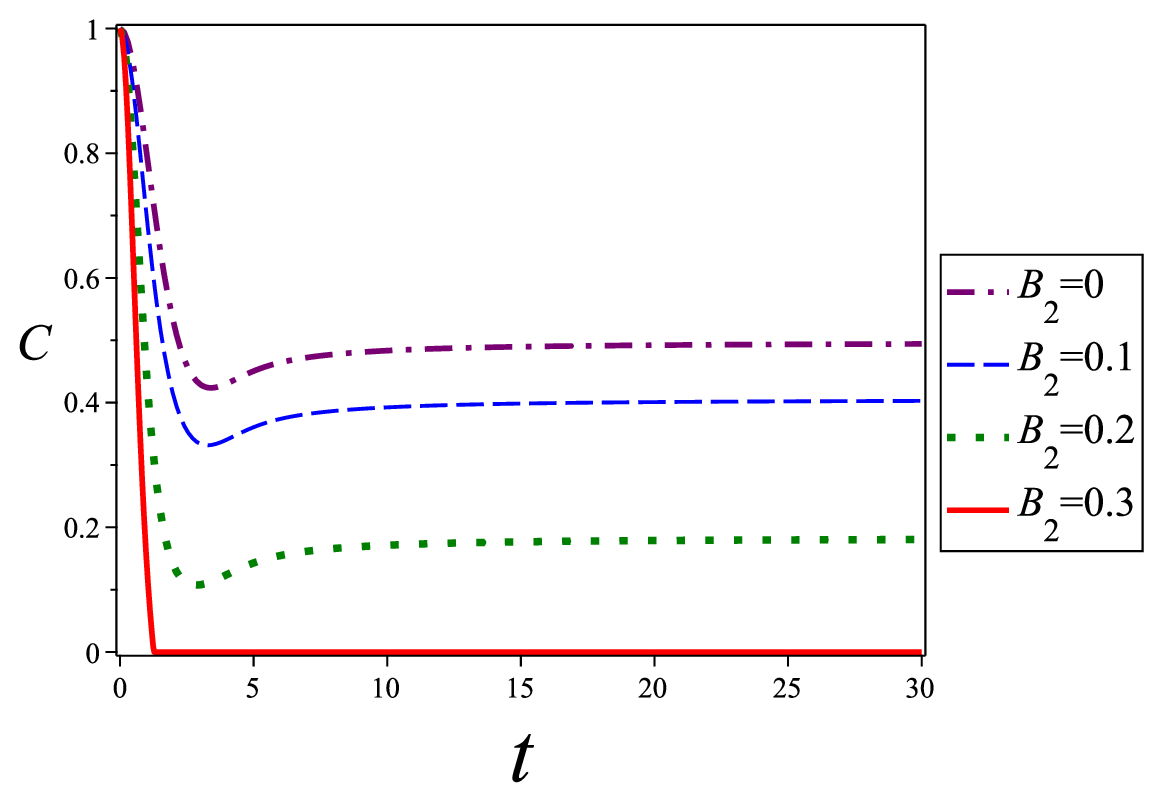}
	\caption{Time variation  of  entanglement, as quantified by concurrence $C(t) $, between the teleported qubits, playing the role of remote probes, for different values of the coupling strength $ B_{2} $ controllable by Bob. }\label{conB2}
\end{figure} 
\begin{figure}[ht!]
	\subfigure[]{\includegraphics[width=9cm]{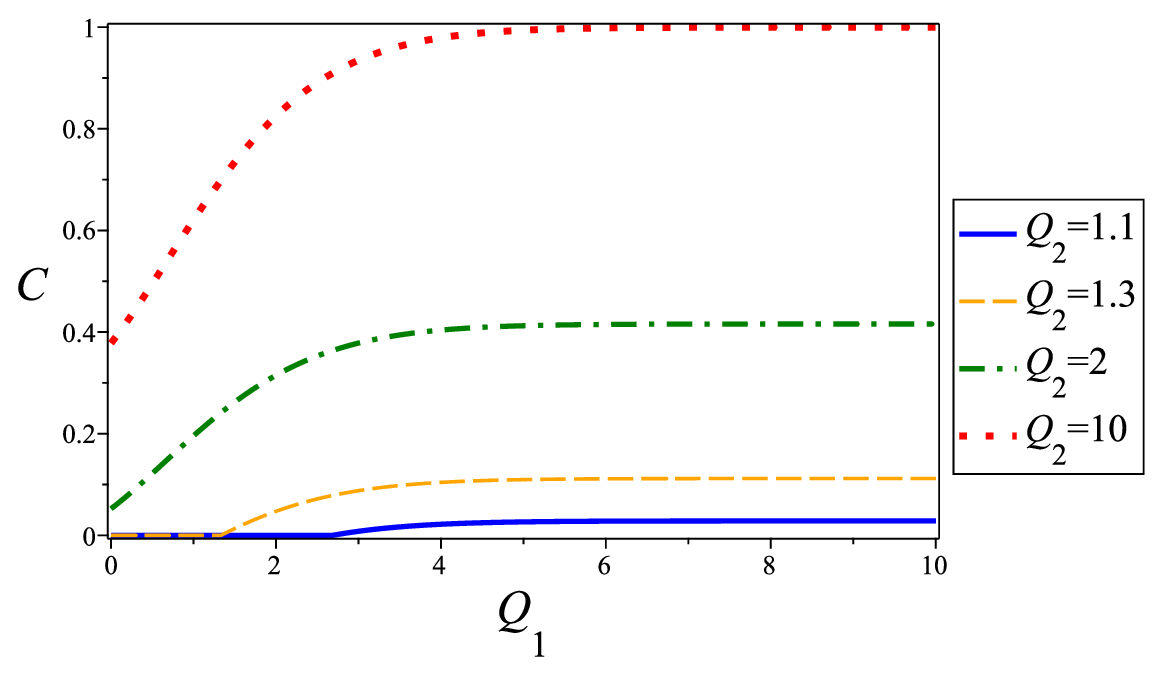}\label{conQ2} }
	\subfigure[]{\includegraphics[width=9cm]{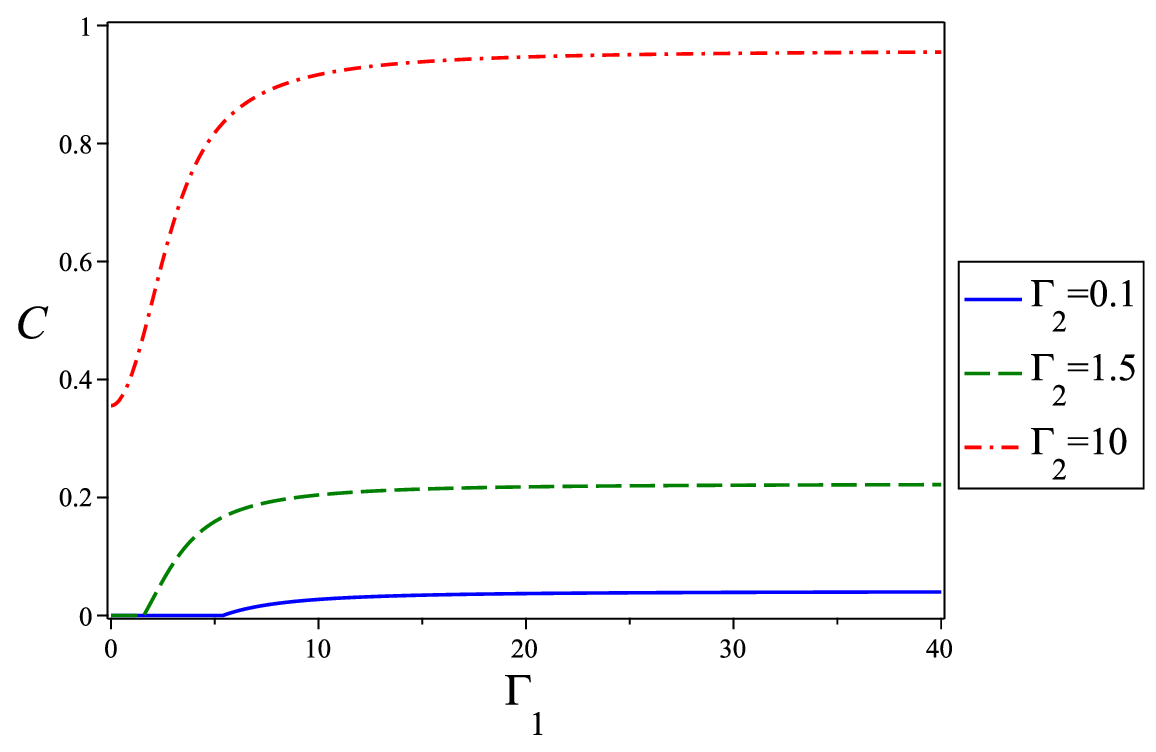}\label{conG2} }
	\caption{(a) Entanglement $ C $ between remote  sensors  as a function of input Ohmicity parameter $ Q_{1} $ for different values of the similar parameter  $ Q_{2} $ controlled by Bob. (b)  The same quantity  as a function of the input cutoff frequency $ \Gamma_{1} $ for different values of $ \Gamma_{2} $ denoting the cutoff frequency controlled by Bob. }
	\label{CQ2G2} 
\end{figure}
It is known that  quantum entanglement between localized qubits, used as quantum probes, may be a resource to achieve advantages in quantum metrology \cite{giovannetti2011advances}. Here we show that an increase in entanglement between the remote probes, realized by non-localized  qubits,   improves the quantum estimation. 
\par
Inserting Eq. (\ref{outputstate}) into Eq. (\ref{a8}), we find that the entanglement of the teleported qubits is give by
\begin{equation}\label{concurrenceout}
C(t)=2\max \left(0,\frac{1}{4} \left(2 \alpha ^2(t)   \sin\theta    \sin^2 \vartheta -\left| \alpha ^4(t)-1\right| \right)\right).
\end{equation}
As seen in Fig. \ref{conB2}, an increase in the intensity of the coupling strength $ B_{2} $, which can be controlled by Bob, decreases the entanglement between the sensors. It is one of the reasons leading to the decay of QFI  with increasing $ B_{2} $, because  entanglement between the teleported qubits is an important resource helping Bob to effectively extract the  information encoded into the sensors.

Moreover,  entanglement may decrease abruptly and non-smoothly to zero in a
finite time due to applying more strong coupling constant. Nevertheless, we find that the QFI is protected from this phenomenon called the \textit{sudden death}.
This shows that there are definitely other resources which can even compensate for the lack of the entanglement of the sensors in the process of remote sensing. 

Because the sudden death of the entanglement may occur even when each of the qubits, used as the teleportation channel, interacts with a non-Markovian environment, we can introduce the non-Markovianty as another resource for the estimation. Nevertheless, protection of the QFI from sudden death in the presence of Markovian environments motivates  more investigation of the  resources playing significant roles in the remote sensing scenario. 
\par
 Figure  \ref{CQ2G2} illustrates the effects of the Ohmicity parameter $ Q_{2} $ and 
cutoff frequency $ \Gamma_{2} $, controlled by Bob implementing the quantum estimation, on   the entanglement of the  sensors and its sudden death. 

 At initial instants, increasing $ Q_{2}~(Q_{1}) $  can improve the entanglement and may decrease the critical value of $ Q_{1}~(Q_{2}) $ at which the \textit{sudden birth} of the entanglement occurs. Therefore, an increase in $ Q_{2}~(Q_{1}) $  can remove the sudden death of the entanglement,  appearing when decreasing $ Q_{1}~(Q_{2}) $ (see Fig. \ref{conQ2} plotted for $ t=0.7 $). It should be noted that
 these considerations are only of relevance to the early moments of the process.
  As time goes on such control is limited and Bob cannot completely counteract  the sudden death of the entanglement by controlling his Ohmicity parameter.
  
  \par
  Investigating the behavior of $ C $ as a function of  $ \Gamma_{1} $ or $  \Gamma_{2} $ for different values of $ Q_{1} $ or $ Q_{2} $, we obtain the similar results. Therefore, the Ohmicity parameters play an important role in controlling  the entanglement of the quantum  sensors and hence their sensitivity for detecting weak coupling coupling constants, because of the positive effect of the probe entanglement on the quantum estimation.   In particular, we see that with large values of both $ Q_{1} $ and $ Q_{2} $ we can completely protect the initial maximal entanglement over time.
  
  Similar control can be implemented by the cutoff frequency as illustrated in Fig. \ref{conG2} plotted for $ t=1.1 $. In detail, studying the behavior of the entanglement between the remote probes as a function of  $ X  \in \left\{Q_{1},Q_{2}, \Gamma_{1}~ (\Gamma_{2})\right\}  $, we see that at initial instants an increase in the cutoff frequency $ \Gamma_{2} ~(\Gamma_{1}) $ can improve the entanglement and may decrease the value of $ X $ at which the sudden birth of the entanglement occurs. In other words, one can remove the sudden death of the entanglement, appearing with decreasing $ X $, with an increase in the cutoff frequency. Again,   as time goes on, such control is limited and we cannot completely counteract  the sudden death of the entanglement by increasing $ \Gamma_{2}~ (\Gamma_{1}) $.

\subsection{Near-perfect teleportation using non-local qubits}
\begin{figure}[ht!]
	\subfigure[]{\includegraphics[width=8.8cm]{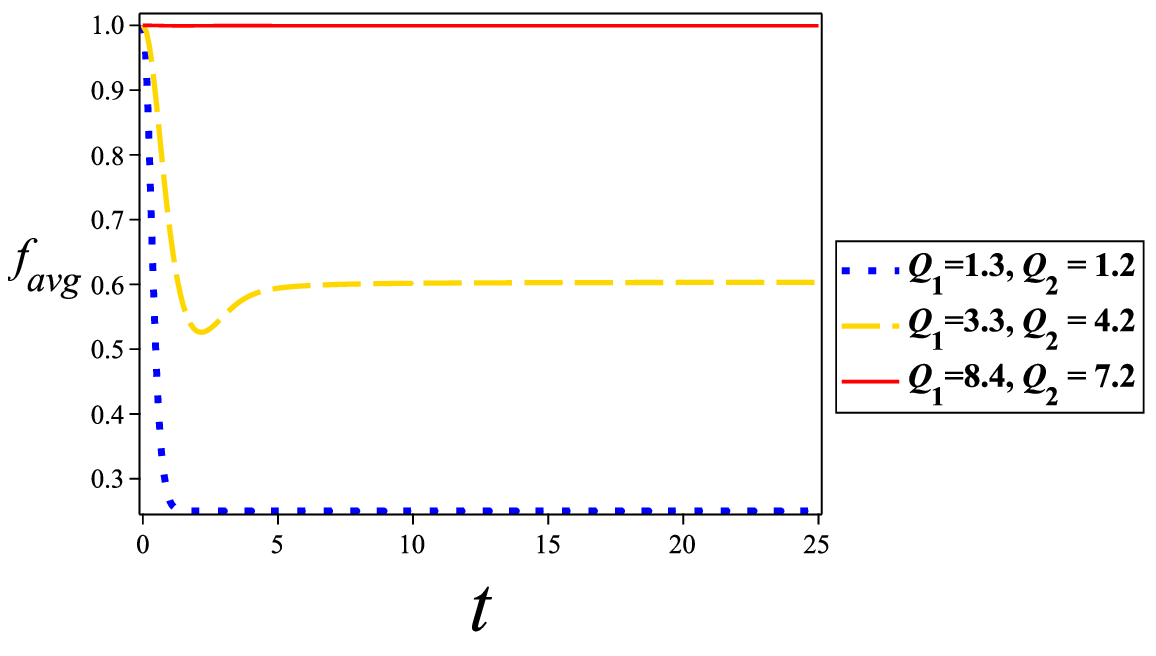}\label{FQ} }
	\subfigure[]{\includegraphics[width=8.8cm]{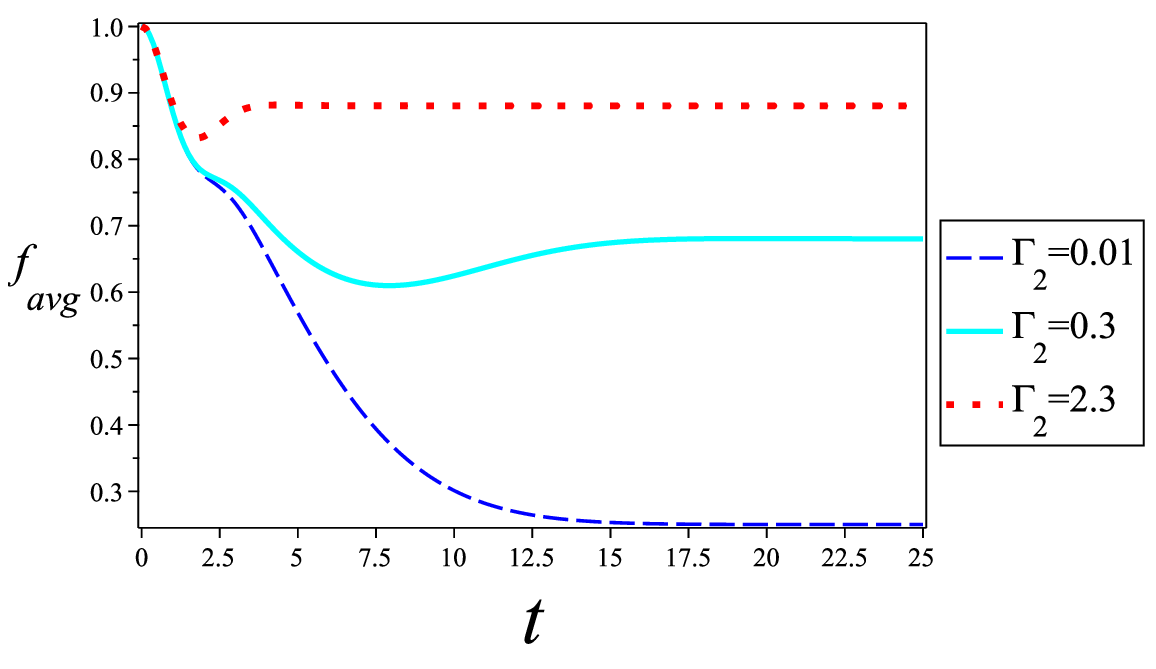}\label{FG} }
	\caption{Dynamics of the average fidelity $ f_{avg} $ of the teleportation  for different values of  (a) Ohmicity parameters $ (Q_{1},Q_{2}) $  and (b)
		cutoff frequency  $ \Gamma_{2} $ controlled by Bob performing the quantum sensing. }
	\label{FQG} 
\end{figure}
\par The quality of teleportation, i.e., closeness of the teleported state to the input state, can be specified by the fidelity \cite{jozsa1994fidelity} between input state $ \rho_{\text{in}}=\ket{\psi_{in}} \bra{\psi_{in}}$ and output state $ \rho_{\text{out}} $. Computing this  measure,  defined as  $  f\left( \rho_{\text{in}},\rho_{\text{out}}\right)=\left\lbrace \text{Tr}\left[\sqrt{\left(\rho_{\text{in}} \right)^{\frac{1}{2}}\rho_{\text{out}}\left( \rho_{\text{in}} \right)^{\frac{1}{2}} }\right]\right\rbrace ^{2}=\langle\psi_{\text{in}}|\rho_{\text{out}}|\psi_{\text{in}}\rangle $. Moreover, the average fidelity, another notion for characterizing the quality of teleportation, is formulated as 
\begin{equation}
	f_{avg}=\frac{\int\limits_{0}^{2\pi}d\varphi\int\limits_{0}^{\pi}f\left( \rho_{\text{in}},\rho_{\text{out}}\right)\sin\theta ~\text{d}\theta}{4\pi}.
\end{equation}
In our model, we find that
\begin{equation}
	f_{avg}=\frac{1}{12} \left(-2 \alpha ^2  \cos2 \vartheta +3 \alpha ^4+4 \alpha ^2+3\right). 
\end{equation}
In classical communication maximum value of average fidelity is given by $ 2/3 $ \cite{popescu1994bell}. Therefore, implementing  teleportation of a quantum state through a quantum resource, with fidelity larger than $ 2/3 $ is worthwhile.

Figure \ref{FQG} shows that how the average fidelity of the teleportation is affected by changing the Ohmicity parameters and cutoff frequency.   When both Ohmicity parameters $ Q_{1} $ and $ Q_{2} $ increase, the average fidelity improves, as shown in Fig. \ref{FQ}. Interestingly, for large values of the Ohmicity parameters,  we can even achieve the quasi-ideal teleportation with $ f_{avg}\approx 1 $. \textit{Therefore, non-local qubits interestingly allows us to    realize near-perfect teleportation with mixed states in the presence of noises. } We emphasize that, according to results presented in \cite{jahromi2019quantum}, this near-perfect teleportation may occur for both Markovian and non-Markovian evolution.

Similar behaviour is observed in terms of   $ \Gamma_{2} $ (see Fig. \ref{FG}).  Bob can increase the quality of teleportation by an increase in his cutoff frequency $ \Gamma_{2} $. In particular, imposing a high cutoff frequency, he can achieve  average fidelity  $f_{avg}> 2/3 $. Therefore,  we can  perform an efficient teleportation
by controlling the cutoff frequency.

\subsection{Comparing output and  channel resources with average fidelity as well as quantum Fisher information}
\begin{figure} [t!]
	\centering
	\includegraphics[width=0.49\textwidth]{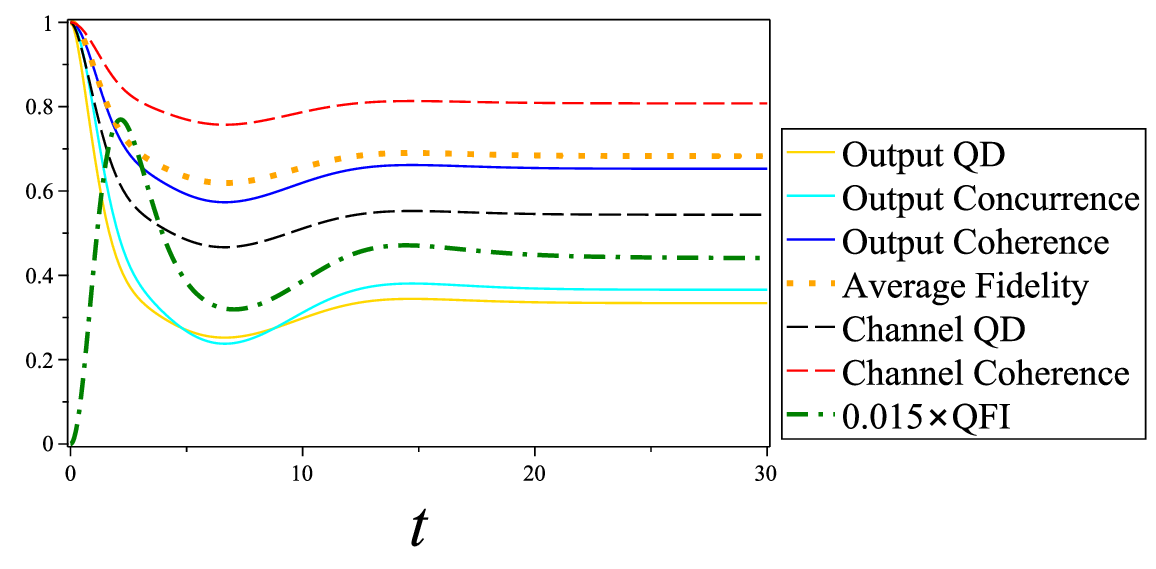}
	\caption{Comparing the dynamics of the average fidelity as well as quantum Fisher information (QFI) with quantum resources (entanglement, quantum discord (QD), and quantum coherence) appearing in the teleportation channel state $ \rho_{ch} $ and those emerging in the output state of the teleportation channel $ \rho_{out} $. }\label{comparison}
\end{figure} 

\par
It is known that  the revivals of
quantum correlations are associated with the non-Markovian
evolution of the system \cite{PhysRevA.85.032318}. Moreover, the previous results \cite{chanda2016delineating, he2017non,radhakrishnan2019time,wu2020detecting} show that  revivals  of quantum coherence can be used for
detecting  non-Markovianity in \textit{incoherent operations} naturally introduced as physical
transformations that do not create coherence. It should be noted that the non-Markovianity, detected by BLP measure of non-Markovianity, discussed in Sec. \ref{Model}, is associated with the violation of \textit{P divisibility} \cite{rivas2010entanglement} and therefore of \textit{CP divisibility} \cite{breuer2009measure} while the non-Markovianity characterized by a coherence measure,  based on a
contractive distance, such as relative entropy \cite{vedral2002role} or $ l_{1} $-norm, corresponds to the violation of  the
CP divisibility. 

Figure \ref{comparison} illustrates the dynamics of quantum resources (quantum entanglement, quantum discord and quantum coherence) of the teleportation output state,  those of the two-qubit system used as the teleportation channel, average fidelity of the teleportation, and the QFI. When the quantum discord and quantum coherence of the
channel revive the non-Markovianity occurs, leading to appearance of the memory effects or  backflow of information from the environment to the two-qubit system applied as the  resource for the quantum teleportation. In the following, we explain the role of this non-Markovian effect to improve the teleportation.

\par
Interestingly, we find that, in the absence of the entanglement sudden death, except for the QFI, all measures exhibit simultaneous
oscillations with time such that their maximum and minimum
points exactly coincide. This excellent agreement among the quantum resources  of the channel, the average fidelity and the teleported quantum resources shows that the non-Markovianity of the channel and  various quantum resources  can be employed  for enhancement of the average fidelity and hence faithfulness of the teleportation. 

\par
Figure \ref{comparison} also shows that comparing the behavior of the QFI with other measures, we cannot faithfully detect the instant at which the optimal  
estimation is achieved. Nevertheless, the time when the QFI is minimized and hence we should avoid it in the process of estimation, can be detected by inspecting the minimum points of the quantum resources.

\section{Conclusions}\label{cunclusion}

Efficient quantum communication among qubits relies on robust networks, which allow for fast and coherent transfer of quantum information.  In this paper we have investigated faithful teleportation of quantum states as well as quantum correlations by non-local topological qubits realized by Majorana modes and independently coupled to non-Markovian Ohmic-like reservoirs.  The roles of control  parameters to enhance the teleportation have been discussed in detail. In Ref. \cite{laine2014nonlocal} the authors showed   that nonlocal memory effects can
substantially increase the fidelity of mixed state quantum teleportation in the presence of dephasing noise such that   perfect quantum teleportation can be achieved even with mixed
photon polarisation states. In their protocol, the nonlocal memory effects occur
due to initial correlations between the local environments of the
photons. Here we have illustrated that using non-local topological qubits, one can perform near-perfect mixed state
teleportation, even in the absence of non-Markovianty and memory effects. 

We have also designed a scheme for remote  sensing through the teleported qubits for estimating some parameter at Alice's location. Without loss of generality, Bob intends to estimate the strength of coupling between Alice's qubits, used as the teleportation resource, and their environments.    It has been shown that the quantum estimation can be considerably enhanced by controlling the environmental parameters whether the evolution is Markovain or  non-Markovianity.
\par
Another important issue which should be addressed is why before the teleportation Bob does not employ his qubits, entangled with Alice's qubits, to estimate $ B_{1} $.  The reason is that by performing local operations, Bob definitely changes the state of the resource channel and hence it may be unusable for quantum  information tasks requiring entanglement such as quantum teleportation, quantum key distribution, etc. Moreover, these changes as well as Bob's activities to estimate some parameter in Alice's location can be detectable by her, while he might want to do it secretly. 

In detail, performing a quantum information task involving communication between Alice and Bob, the two parties should consider two important security conditions \cite{zhu2006secure,li2006improving,chou2018dynamic,chou2021multiparty}: security against  internal and external attacks. In internal attacks, either Alice or Bob attempts to steal the other’s secret information. However, in external attacks  an eavesdropper, Eve, attempts to steal messages without being detected by Alice or Bob. Because of these security considerations, always performing security checks by Alice and Bob is always necessary to detect internal or external attacks. Bob's activities to estimate one of Alice's parameters through the qubits used for the resource channel can be detected by Alice in the security check.  Therefore, Bob prefers to hide his activities from Alice and employ teleported qubits for implementing the remote sensing.

\par
Recently,   a quantum simulation
of the teleportation of a qubit encoded as the Majorana
zero mode states
of the Kitaev chain has been performed \cite{huang2021emulating}. Our work motivates further studies on physical realization of quantum teleportation by large number of non-local topological qubits and paves the way for designing remote sensors based on Majorana modes.

\section*{Declaration of competing interest}
We have no competing interests.

\section*{Acknowledgements}
  I  wish to acknowledge the financial support of the MSRT of Iran and Jahrom University.
 I am very grateful to  Mostafa Rostampour and Fazileh Aminizadeh for helping me in designing  Fig. 1.

\bibliography{Ref}

\end{document}